\documentclass[aps,twocolumn,groupedaddress,prb,floatfix,showpacs]{revtex4-1}
\usepackage{mathrsfs}
\usepackage{amsmath}
\usepackage[dvips]{graphics,color}
\usepackage{graphicx}
\usepackage{epstopdf}
\usepackage{dcolumn}
\usepackage{bm}
\usepackage{longtable}
\usepackage{graphics}
\usepackage{amssymb}
\usepackage{xspace}
\usepackage{epsfig}
\usepackage{dcolumn}
\usepackage[FIGTOPCAP,hang,raggedright,nooneline,bf]{subfigure}
\begin{document}
\title{Competing Ordered States in Bilayer Graphene}
\author{Fan Zhang$^1$}
\email{zhf@sas.upenn.edu}
\author{Hongki Min$^2$}
\author{A.H. MacDonald$^3$}
\affiliation{
$^1$ Department of Physics and Astronomy, University of Pennsylvania, Philadelphia, PA 19104, USA\\
$^2$ Department of Physics and Astronomy, Seoul National University, Seoul 151-747, Korea\\
$^3$ Department of Physics, University of Texas at Austin, Austin TX 78712, USA
}
\date{\today}

\begin{abstract}
We use a perturbative renormalization group approach with short-range continuum model interactions
to analyze the competition between isotropic gapped and anisotropic gapless ordered states in bilayer graphene,
commenting specifically on the role of exchange and on the importance of spin and valley flavor degeneracy.
By comparing the divergences of the corresponding susceptibilities, we
conclude that this approach predicts gapped states for flavor numbers $N=1,2,4$.
We also comment briefly on the related gapped states expected in chiral (ABC) trilayer graphene.
\end{abstract}

 \maketitle

\section{Introduction}

In chirally stacked few layer graphene systems \cite{graphene_review,AB_McCann,Chiral_Min,ABC_Zhang,SQH_Zhang} two sublattice sites, one in the
top layer and one in the bottom layer, are disconnected from the near-neighbor interlayer hopping network
and partially isolated.  This geometric feature leads to a gapless semiconductor
with weakly dispersive conduction and valence band edges,
and therefore to an enhanced role for electron-electron interactions \cite{ABC_Zhang,SQH_Zhang}.  The low-energy sublattice degree-of-freedom in chirally stacked $N$ layer graphene can be viewed as a $S=1/2$ pseudospin, with $\uparrow$-pseudospins corresponding to top
layer states and $\downarrow$-pseudospins to bottom layer states.  In this language the conduction and valence band
eigenstates at wavevector $\bm{q}=q\,(\cos\phi_{\bm{q}},\sin\phi_{\bm{q}})$ have $\hat{x}-\hat{y}$ plane pseudospin orientations, {\em i.e.} they have equal weight in top and bottom layers, and azimuthal angles $N \phi_{\bm{q}}+\pi$ and $N \phi_{\bm{q}}$ respectively.
A number of years ago we predicted\cite{MF_Min} on the basis of an unrestricted Hartree-Fock
mean-field-theory that these circumstances would lead to ground states with pseudospin order, and in particular
to a state in which the pseudospins rotate out of the $\hat{x}-\hat{y}$ plane by a $k$-dependent
polar angle toward either the north or the south pole, breaking layer-inversion symmetry \cite{ft_inversion}
and inducing a gap \cite{RG_Zhang} in the charged excitation spectrum.
Recent bilayer experimental work \cite{Yacoby1,Yacoby2,Geim,AB_Schonenberger,AB_Lau,AB_Lau1,AB_Wees}
has yielded clear evidence for broken symmetry states, although a consensus on the character of their
order has not yet been achieved.  At the same time a substantial body of theoretical
work \cite{Fradkin,caf1,caf2,MF_Levitov,RG_Vafek,RG_Zhang,RG_Levitov,RG_Falko,Trushin_1,MF_MacDonald,
SQH_Zhang,MF_Jung,SQH1_Zhang,LAF_Zhang,SQH_Levitov,Vafek_2,Vafek_3,Falko_2,Trushin_2,fRG,Varma,Dagim,prepare,GW}
has used a variety of different approaches to study bilayer graphene.  The general consensus of this
work is that the ground state is a pseudospin ferromagnet.  It has become clear however that because of
quantum fluctuations that are not captured by mean-field theory, the gapped state competes closely with
a gapless anisotropic state in which the net pseudospin alignment is in the $\hat{x}-\hat{y}$ plane, not
the $\hat{z}$ direction.  This paper addresses that competition.

As we will explain in more detail later, some features of the bilayer graphene system present awkward
obstacles to theory.  First among these is the property that the conduction and valence bands are weakly
dispersive only over the small part of the Brillouin zone where inter-layer tunneling plays an essential role.
Elsewhere in the Brillouin zone interaction effects are expected to be similar to those in
single-layer graphene, producing strong quasiparticle
velocity enhancements,\cite{Elias_Velocity,Paco_Velocity,Borghi_Velocity}
and spectral weight rearrangements,\cite{plasmaron_expt,plasmaron_theory} but not order.
The small number ($\sim 10^{-4}$) of $\pi$-electrons per carbon atom in the strongly correlated region of momentum space is an obstacle
for theories of the ordered state that are
based primarily on a lattice model of graphene.  The narrow momentum-space
distribution of the most important virtual states also creates difficulties for continuum models
because it makes the physics sensitive to the long-range of the Coulomb interaction.
In this paper we report on calculations in which lattice effects are completely ignored
and the long-range of the Coulomb interaction is not treated explicitly.  In our view the
continuum approach we use is strongly motivated by the low-density of strongly correlated electrons.
Our use of short-range interactions can be crudely justified by appealing to screening considerations,
and arguing that the momentum-independent interaction parameters we use represent an
average over the relevant portion of momentum space.
Although not fully rigorous in this sense, we believe that the
conclusions we reach by following this approach systematically are nevertheless valuable
and consistent with recent experimental work.\cite{Yacoby1,Yacoby2,Geim,AB_Schonenberger,AB_Lau,AB_Lau1,ABC_Lau,AB_Wees}

This paper elaborates on ideas presented in a series of
closely related earlier papers \cite{RG_Vafek,RG_Zhang,RG_Levitov,RG_Falko,fRG}
in which the ground state is estimated on the basis of perturbative renormalization group (PRG) calculations.
These calculations differ among themselves most essentially in the
approximate ways they account for lattice scale physics and long-range
interactions.  In Section II we briefly describe the model we study and restate some
identities and observations that are useful in developing its perturbation theory.
Section III presents new results for the PRG interaction flows of this model, emphasizing the
role of exchange interactions between electrons with the same spin and valley states and
commenting on the flavor number ($N$) dependence of the interaction parameter flows.
In Section IV we explain how the PRG pseudospin susceptibilities, whose
divergences suggest the character of the low-temperature order, depend on the renormalized interaction strengths.
Finally in Section V we present our conclusions and discuss the relationship of our work to
experiments and to other theoretical work on this system.

\section{Theoretical Preliminaries}

Bilayer graphene (BLG) is described approximately at low-energies by the $J=2$ version
of the chiral band Hamiltonian:\cite{AB_McCann,Chiral_Min,ABC_Zhang,SQH_Zhang}
\begin{equation}
\label{eq:band} {\mathcal H}_{\rm J}=\sum_{{\bm q}\alpha\beta i}
\frac{(\hbar v q)^{\rm J}}{(-\gamma_1)^{\rm J-1}}c^{\dagger}_{{\bm q}\alpha i}\big[\cos(J
\phi_{\bm q})\sigma^{x}_{\alpha\beta}+\sin(J\phi_{\bm q})
\sigma^{y}_{\alpha\beta}\big]c_{{\bm q}\beta i}.
\end{equation}
In Eq.~(\ref{eq:band}) ${\bm \sigma}$ are Pauli matrices which
act on the (Greek) layer labels and $i=1,\ldots,N$ is a spin-valley flavor label.
We have used the notation $\cos\phi_{\bm q}=\tau^{\rm z}q_{\rm x}/q$ and $\sin\phi_{\bm q}=q_{\rm y}/q$. Here $\tau^{\rm z}=\pm 1$
labels $K$ and $K'$ valleys located at inequivalent Brillouin zone (BZ) corners and
${\bm q} = (q_x,q_y)$ is wavevector measured from the BZ corner.
It is convenient to perform a rotation in BLG pseudospin space\cite{ft_valley} in the $K' (\tau_{z}=-1)$ valley
to eliminate the valley dependence of the band Hamiltonian, in order to make our discussion more concise in the following sections. The BLG chiral band Hamiltonian applies \cite{AB_McCann,Chiral_Min} at energies
smaller than the interlayer hopping scale
$\gamma_1 \sim 400$ meV and larger than the trigonal warping
scale $\sim 1$ meV.  We discuss the important role of trigonal warping in the band-structure later.
The physics discussed in this paper applies to bilayers when $J=2$ and partly
generalizes to $J$-layer chirally (ABC) stacked
graphene multilayers,\cite{ABC_Lau,ABC_Zhang,ABC_McCann,3QHE_Zhang}
although the chiral band Hamiltonian ${\mathcal H}_{\rm J}$ applies
over narrower ranges of energy for larger $J$.
We will focus our attention on the $J=2$ case hereafter.

${\mathcal H}_{2}$ can be viewed as specifying a
momentum ${\bm q}$-dependent effective magnetic field that
acts on the bilayer layer-pseudospin degree-of-freedom:
\begin{equation}
{\mathcal H}_2= - {\bm B}\cdot {\bm \sigma}~,
\end{equation}
where $|{\bm B}|=\xi_{\bm q}=\hbar \omega_{\bm q} = \hbar^2q^2/2m^*$ with $m^*=\gamma_1/(2v^2)$, and the
orientation angle of ${\bm B}$ is $2 \phi_{\bm q}$.
The corresponding Matsubara Green's function is
\begin{equation}
{\mathcal G}_{\bm q}(i\omega_{\rm n})= [i\omega_{\rm n}- {\mathcal
H}/\hbar]^{-1} = {\mathcal G}_{{\bm q}\rm s}(i\omega_{\rm n})+{\mathcal
G}_{{\bm q}\rm t}(i\omega_{\rm n}) \; {\bm n} \cdot {\bm \sigma}\,
\end{equation}
where
\begin{equation}
{\mathcal G}_{{\bm q}\rm s,t}(i\omega_{\rm n})\equiv{1\over
2}\left({1\over i\omega_{\rm n}-\omega_{\bm q}}\pm{1\over
i\omega_{\rm n}+\omega_{\bm q}}\right),
\end{equation}
and ${\bm n} = -\,(\cos2\phi_{\bm q},\sin2\phi_{\bm q},0)$.
Note that ${\mathcal G}_{\rm
s}(-i\omega_{\rm n}) = -{\mathcal G}_{\rm s}(i\omega_{\rm n})$
whereas ${\mathcal G}_{\rm t}(-i\omega_{\rm n}) = {\mathcal G}_{\rm
t}(i\omega_{\rm n})$.  When expressed explicitly as a $2 \times 2$ matrix,
\begin{equation}
\label{eq:phasefactor} {\mathcal G}(i\omega_{\rm n})= \left(
\begin{array}{cc}
{\mathcal G}_{\rm s}(i\omega_{\rm n}) & -{\mathcal G}_{\rm t}(i\omega_{\rm n}) e^{\rm -2 i \phi_{q}} \\
-{\mathcal G}_{\rm t}(i\omega_{\rm n}) e^{\rm 2 i \phi_{q}} & {\mathcal G}_{\rm s}(i\omega_{\rm n})  \\
\end{array}
\right).
\end{equation}
The off-diagonal {\em triplet} component captures processes in which
electrons propagate between layers; its momentum-orientation
dependent phase factor plays an essential role in the one-loop PRG calculations
described in the next section.
The frequency sums which appear in loop diagrams are readily evaluated:
\begin{eqnarray}
\label{eq:freq_sum_gf} &&{1\over \beta\hbar^2}\sum_{\omega_{\rm n}}
{\mathcal G}_{{\bm q}{\rm s}, {\rm t}}^2(i\omega_{\rm n}) = \mp{\tanh
(\beta \xi_{\bm q}/2) \over 4 \xi_{\bm q} }\mathop
{\longrightarrow} \limits_{T \to 0} \mp{1\over 4 \xi_{\bm q}}\,;\nonumber\\
&&{1\over \beta\hbar^2}\sum_{\omega_{\rm n}} {\mathcal G}_{{\bm q}\rm
s}(i\omega_{\rm n}) {\mathcal G}_{{\bm q}\rm t}(i\omega_{\rm n}) \mathop
{\longrightarrow} \limits_{T \to 0} 0\,.
\end{eqnarray}
These two results express the property that fluctuations involve interband
transitions.  In the PRG each loop diagram
is multiplied by appropriate interaction constants (discussed below)
and then integrated over momentum labels
near the model's flowing cutoff $\Lambda$:
\begin{equation}
\int_{\Lambda/ s <q<\Lambda} \frac{d^2 {\bm q}}{(2\pi)^2} \;
\frac{\tanh(\beta \xi_{\bm q}/2)}{4\xi_{\bm q}} \; \mathop
{\longrightarrow} \limits_{T \to 0} \;\frac{1}{2}  {\nu_0} \, \ln(s).
\end{equation}
where $\nu_0=m^*/2\pi\hbar^2$ is the BLG
density-of-states per flavor.
The second frequency sum in Eq.(\ref{eq:freq_sum_gf}) vanishes in the long-wavelength
static limit because it has contributions from
intra-band particle-hole excitations only.
Because $\omega_{\bm q} \propto q^2$ rather than
$q$, this integral diverges logarithmically when the high-energy cut-off is
scaled down by a factor of $s$.  The PRG calculation is therefore very similar to the corresponding
calculation for a 1DES.  This rather surprising property of
BLG is directly related to its unusual band structure which yields
Fermi points rather than Fermi lines in neutral systems, and quadratic rather than
linear dispersion.\cite{RG_Zhang}

\section{RG Flow Equations}

\subsection{Number of Physical Parameters}
In systems with short-range interactions, finite
momentum (gradient) corrections are irrelevant in the RG flows.
We therefore ignore the wave-vector
transfer dependence of scattering processes starting
already at the $J=2$ chiral model's ultraviolet cut-off energy $\sim \gamma_1$.
In the low-energy continuum model of BLG electrons
carry spin, and both layer and valley pseudospin labels. In a
scattering event, both the two incoming and two outgoing particles
can therefore have one of eight labels and the general scattering
function therefore has $8^4$ distinct
values, even when gradient corrections are ignored.
The number of distinct coupling constants in the RG flow
equations is much smaller, however, because many values are zero and
others are related to each other by symmetry. One simplification is
that interactions conserve spin and, in the continuum model, both layer and valley
pseudospin at each vertex.  The role of spin in our model is therefore equivalent to the role of
valley.   We do however choose to allow the interaction strength between
electrons in different ($D$) layers to be weaker than that between electrons in the same ($S$)
layer so that the interactions are layer-pseudospin dependent.

\begin{figure}[htbp]
\centering \scalebox{0.5} {\includegraphics*
[1.9in,7.3in][8.2in,9.55in] {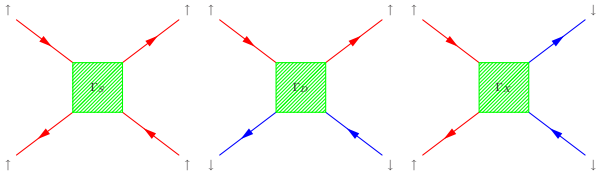}}
\caption{\label{fig:gamma}(Color online) Electron-electron
scattering processes for a system with a single pseudospin-$1/2$ degree
of freedom.}
\end{figure}

The internal loops in the PRG
calculation contain two fermion propagator (Green's function) lines.
Because we seek a momentum-independent effective interaction we
simplify the calculation by evaluating the coarse-graining correction to the effective interaction
for the case where all incoming and outgoing electrons are at the Dirac point.
(As we comment in the discussion section, this
approximation is made less innocent by the layer pseudospin structure of the band eigenstates.)
These propagators conserve both spin and valley pseudospin, but as
we have seen above, not the layer pseudospin.  From Eq.(\ref{eq:phasefactor})
we see that a phase factor \cite{ft_valley} $e^{\rm \pm 2i\phi_{q}}$ appears when a propagator transfers electrons
between layers.  Unless these layer transfers enter an equal number
of times in each direction, the integrand in a Feynman diagram will
contain a net phase factor related to chirality and vanish upon angular integration of the virtual
momentum. The total layer number (the $\hat{z}$ component of layer pseudospin) is therefore also
conserved.\cite{ft_valley}  This property combined with flavor invariance limits the number of independent interaction
parameters to three, two for interaction processes involving electrons in the same layer
or different layers that conserve layer index, and one for interaction processes between electrons
in different layers accompanied which interchange layer indices.  We refer to the three distinct processes,
illustrated in Fig.~\ref{fig:gamma},
as $\Gamma_{S}$, $\Gamma_{D}$ and $\Gamma_{X}$ respectively.

If flavor invariance is broken, either due to microscopic lattice effects that violate $SU(4)$ invariance
or due to broken symmetries, the number of interaction parameters increases.
For example, if the bare interactions are spin and valley dependent in
an arbitrary way, the number of independent {\em antisymmetrized} interaction parameters increases from three to ten:\cite{RG_Zhang}
$\Gamma_{\rm SSD}$,$\Gamma_{\rm SDS}$, $\Gamma_{\rm SDD}$,
$\Gamma_{\rm DSS}$,$\Gamma_{\rm DSD}$, $\Gamma_{\rm DDS}$, $\Gamma_{\rm DDD}$,
$\Gamma_{\rm XSD}$, $\Gamma_{\rm XDS}$ and $\Gamma_{\rm XDD}$.
In this notation the first label follows the convention explained above for the layer pseudospin
dependence of the interaction, whereas the second and third labels refer to
valley and spin respectively, and distinguish the interaction between particles with the
same ($S$) or different ($D$) spin or valley.
The number of entries in this list of interactions has been reduced by two by appealing to Fermi
statistics for the case of $SS$ interactions, those between particles with the same spin and valley labels.
In that case interchanging the labels of outgoing electrons yields an indistinguishable and canceling
amplitude for particles in the same layer.
Although we can choose to retain $\Gamma_{\rm SSS}$ in the RG flows, it
cannot contribute to any physical observable.  Similarly, since $\Gamma_{\rm DSS}$ and $\Gamma_{\rm XSS}$
are interchanged by reversing outgoing labels they differ only by a sign and it is sufficient to retain one parameter.
The number of independent interacting parameters in the analysis by Vafek \cite{Vafek_2} and by Lemonik {\em et al.},\cite{Falko_2} who impose the symmetry of the underlying BLG crystal but do not explicitly
account for antisymmetrization is nine.
For $SU(4)$ invariant models the number of parameters is three, but
antisymmetrization allows us to retain only one coupling constant in the single flavor ($N=1$) case,
and the physical implications of the flowing interactions are more apparent if we do so.\cite{RG_Zhang}
A convenient choice \cite{RG_Zhang} is to retain $\Gamma_{\rm DSS}$.
We take the view that the underlying lattice \cite{MF_Jung} is unlikely to play an
important role in determining whether the broken symmetry state is gapped \cite{RG_Zhang} or
gapless,\cite{RG_Vafek,RG_Falko} though it is key to selecting \cite{MF_Jung} between distinct gapped states \cite{SQH_Zhang}
that are equivalent in an $SU(4)$ invariant model.   Therefore,
the number of interactions parameters that appear in the calculation described below is
three in the general case with unbroken flavor invariance and one for the special case of $N=1$.

\subsection{RG Flow for $N=1$}

In this subsection we explore the similarities and differences between
BLG and 1DES's by temporarily neglecting the spin and valley degrees-of-freedom.
We choose the single effective interaction parameter to be $\Gamma_{\rm D}$.
A PRG analysis determines how the bare interaction at the $\gamma_1$ scale,
$V_{\rm D}$, is renormalized by
integrating out the high energy fermion degrees of freedom.
At one loop level the coarse-graining contributions to the effective
interaction are described \cite{RGShankar} by the three higher order
diagrams labeled ZS, ZS', and BCS.  The internal loops in these diagrams are summed over the high-energy
labels.  The main merit of the PRG is that it treats all virtual processes on an equal footing,

It is well known that in a 1DES the ZS loop vanishes while the ZS'
and BCS diagrams cancel, implying that the interaction strength does
not flow to large values and hence that neither the CDW repulsive interaction
nor the BCS attractive interaction instabilities
predicted by mean-field theories survive the quantum fluctuations
they neglect.  The key features of BLG order physics can be understood
in terms of two properties of these one-loop diagrams; (i) the particle-particle
(BCS) and particle-hole (ZS, ZS') loops have the same logarithmic
divergences as in the 1DES case in spite of the larger space
dimension and (ii) the ZS loop, which vanishes in the 1DES case, is
finite in the BLG case and the BCS loop vanishes
instead.  Both of these changes are due to the layer pseudospin
triplet contribution to the single-particle Green's function as we
explain below.  The net result is that interactions flow to strong
coupling even more strongly than in the mean-field approximation.

The key step in one-loop PRG calculations is identifying the
coupling factors attached to the loop in each diagram.
Since only opposite layer
interactions need to be retained for $N=1$ models, all scattering functions have two
incoming particles with opposite layer labels and two outgoing
particles with opposite layer labels, much like the 1DES case.\cite{Giamarchi}
The external legs in the
scattering function Feynman diagrams are
labeled by the layer index ($T =$ top layer and $B = $ bottom layer).
The corresponding labels for the 1DES case \cite{RGShankar,Giamarchi} are the chirality, $R$ for
right-going and $L $ for left-going.  We refer to these as the single-particle labels below
when a comment refers to both 1DES and BLG cases.
\begin{figure}[htbp]
\subfigure[\color{white}{a}]{\scalebox{0.63}{\includegraphics*[2.1in,7.3in][4.4in,9.55in]{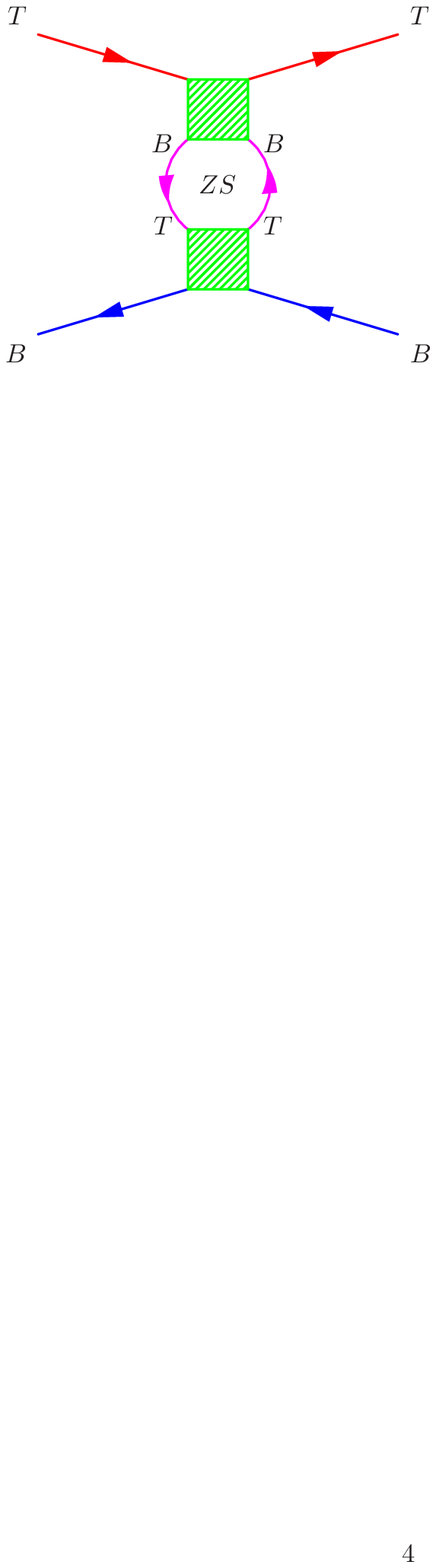}}}\hspace{0.4in}
\subfigure[\color{white}{a}]{\scalebox{0.63}{\includegraphics*[2.1in,7.3in][4.4in,9.55in]{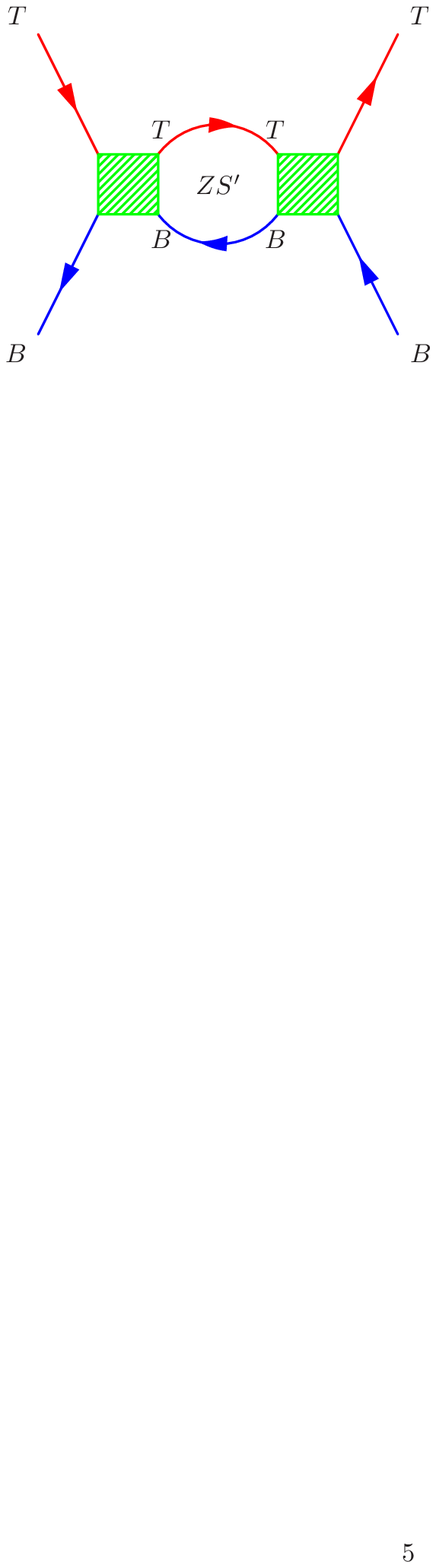}}}
\caption{\label{fig:zszs1}(Color online) {\bf (a)} ZS, {\bf (b)} ZS'
loop corrections in the one-loop PRG calculation.}
\end{figure}

As shown in Fig.~\ref{fig:zszs1}(a), at the upper vertex of the ZS
diagram the incoming and the outgoing $T$ particles induce a $B$
particle-hole pair in the loop, while the incoming and outgoing $B$
particles at the lower vertex induce a $T$ particle-hole pair.
The ZS contribution is absent in the 1DES
case \cite{RGShankar,Giamarchi} because propagation is always
diagonal in interaction labels. However, this correction survives
for BLG because the single-particle Green's function has a triplet
contribution (see Eq.~(\ref{eq:freq_sum_gf})) which is off-diagonal
in layer index.  Here we find
\begin{eqnarray}
\label{eq:ZS}
\Gamma_{\rm D}^{\rm ZS}={\Gamma_{\rm D}^2\over \beta\hbar^2}\int
{d^2 {\bm q}\over (2\pi)^2}\,\sum_{\omega_{\rm n}}
{\mathcal{G}}_{\rm t}^2(\bm{q},i\omega_{\rm n})={1\over 2}\, \Gamma_{\rm
D}^2\,\nu_0\ln (s)\,.
\end{eqnarray}

The ZS' loop shown in Fig.~\ref{fig:zszs1}(b) corresponds to repeated
interaction between a $T$ particle and a $B$ hole. This is the
channel responsible for the 1DES mean-field CDW
instability \cite{RGShankar} in which coherence is established
between $R$ and $L$ particles.\cite{Giamarchi} In both 1DES and BLG
cases it has the effect of enhancing repulsive interactions. Its
evaluations for the two cases correspond quite closely, because this
loop diagram involves only particle-propagation that is diagonal in
interaction label.  We find that
\begin{eqnarray}
\Gamma_{\rm D}^{{\rm ZS'}}=-{\Gamma_{\rm D}^2\over
\beta\hbar^2}\int {d^2 {\bm q}\over (2\pi)^2}\,\sum_{\omega_{\rm n}}
{\mathcal{G}}_{\rm s}^2(\bm{q},i\omega_{\rm n})={1\over 2}\, \Gamma_{\rm
D}^2\,\nu_0\ln (s)\,.
\end{eqnarray}
\begin{figure}[htbp]
\centering \scalebox{0.63}
{\includegraphics*[2.1in,7.3in][6.9in,9.55in]{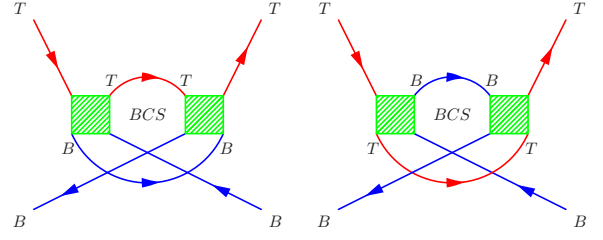}}
\caption{\label{fig:bcs1}(Color online) BCS (particle-particle) loop
correction for singlet propagation in the one-loop PRG calculation.}
\end{figure}

The BCS channel corresponds to repeated interaction between the two
incoming particles. In the 1DES case the contribution from this loop
(see Fig.~\ref{fig:bcs1}) cancels the ZS'
contribution\cite{RGShankar}, leading to marginal interactions and
Luttinger liquid behavior. This same kind of BCS correction for
BLG reads
\begin{eqnarray}
\label{eq:bcs1} \Gamma_{\rm D}^{{\rm BCS}_1}=&-&{1\over
2}{\Gamma_{\rm D}^2\over \beta\hbar^2}\int {d^2 {\bm q}\over
(2\pi)^2}\,\sum_{\omega_{\rm n}}
{\mathcal{G}}_{\rm s}(\bm{q},i\omega_{\rm n}){\mathcal{G}}_{\rm s}(-\bm{q},-i\omega_{\rm n})\nonumber\\
&-&{1\over 2}{(-\Gamma_{\rm D})^2\over \beta\hbar^2}\int {d^2 {\bm
q}\over
(2\pi)^2}\,\sum_{\omega_{\rm n}} {\mathcal{G}}_{\rm s}(\bm{q},i\omega_{\rm n}){\mathcal{G}}_{\rm s}(-\bm{q},-i\omega_{\rm n})\nonumber\\
=&-&{1\over 2}\Gamma_{\rm D}^2\,\nu_0\ln (s)\,.
\end{eqnarray}
\begin{figure}[htbp]
\centering \scalebox{0.63} {\includegraphics*
[2.1in,7.3in][6.9in,9.55in] {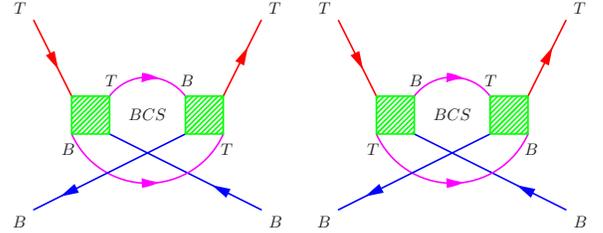}}
\caption{\label{fig:bcs2}(Color online) BCS (particle-particle) loop
correction for triplet propagation in the one-loop PRG calculation.}
\end{figure}

In the BLG case there is an additional BCS loop
contribution (see Fig.~\ref{fig:bcs2})
in which the incoming $T$ and $B$ particles both change layer labels
before the second interaction.  This contribution is possible
because of the triplet pseudospin propagation and, in light of
Eq.~(\ref{eq:freq_sum_gf}), gives a BCS contribution that has the opposite
sign compared to the normal contribution:
\begin{eqnarray}
\Gamma_{\rm D}^{{\rm BCS}_2}&=&-{1\over 2}{\Gamma_{\rm
D}(-\Gamma_{\rm D})\over \beta\hbar^2}\int {d^2 {\bm q}\over
(2\pi)^2}\,\sum_{\omega_{\rm n}}
{\mathcal{G}}_{\rm t}(\bm{q},i\omega_{\rm n}){\mathcal{G}}_{\rm t}(-\bm{q},-i\omega_{\rm n})\nonumber\\
&&-{1\over 2}{(-\Gamma_{\rm D})\Gamma_{\rm D}\over \beta\hbar^2}\int
{d^2 {\bm q}\over (2\pi)^2}\,\sum_{\omega_{\rm n}}
{\mathcal{G}}_{\rm t}(\bm{q},i\omega_{\rm n}){\mathcal{G}}_{\rm t}(-\bm{q},-i\omega_{\rm n})\nonumber\\
&=&{1\over 2}\Gamma_{\rm D}^2\,\nu_0\ln (s) \,.
\end{eqnarray}

It follows that the BCS loop contribution is absent in the BLG case because
\begin{eqnarray}
\Gamma_{\rm D}^{\rm BCS}=\Gamma_{\rm D}^{{\rm BCS}_1}+\Gamma_{\rm
D}^{{\rm BCS}_2}=0\,.
\end{eqnarray}
Therefore, at one-loop level, the renormalization of interlayer
interaction is
\begin{eqnarray}
\Gamma_{\rm D}^{{\rm one-loop}}=\Gamma_{\rm D}^{\rm ZS}+\Gamma_{\rm
D}^{{\rm ZS'}}+\Gamma_{\rm D}^{\rm BCS}=\Gamma_D^2\,\nu_0\ln(s)\,.
\end{eqnarray}
These results and the comparison with 1DES's are summarized in
Table~\ref{table:one} and imply the following RG flow equation for
BLG:
\begin{eqnarray}
\frac{{\emph d}\,\Gamma_{{\rm D}}}{\nu_0\,{\emph
d}\ln(s)}=\Gamma_{\rm D}^2\,.
\end{eqnarray}
If we follow the flows from the model's microscopic cutoff
we obtain
\begin{equation}
\Gamma_{\rm D} = \frac{V_{\rm D}}{1-V_{\rm D}\nu_0 \ln(s)}
\end{equation}
where $V_{\rm D}$ is the bare interaction.  $\Gamma_{\rm D}$
diverges when $V_{\rm D}\,\nu_0 \geq 1/ \ln(s)$. This
instability criterion is similar to the Stoner criterion for
ferromagnetism.
\begin{table}[t]
\caption{Summary of contributions from the
three one-loop diagrams to $\Gamma_{\rm D}$ in units of $\nu_{0}$
in 1DES and BLG cases.} \centering
\newcommand\T{\rule{0pt}{3.1ex}}
\newcommand\B{\rule[-1.7ex]{0pt}{0pt}}
\begin{tabular}{c  c  c }
\hline\hline diagrams &\qquad \qquad 1DEG\qquad\T &\qquad BLG\\[3pt]
\hline{\em ZS}  &\qquad $0$\T &\qquad $\frac{1}{2}\,\Gamma_{\rm D}^2\,\ln(s)$ \\[3pt]
{\em ZS'} &\qquad $u^2\,\ln(s)$\T &\qquad $\frac{1}{2}\,\Gamma_{\rm D}^2\,\ln(s)$ \\[3pt]
{\em BCS} &\qquad $-u^2\,\ln(s)$\T &\qquad $0$ \\[3pt]
\hline Mean Field &\qquad $u^2\,\ln(s)\T$ &\qquad $\frac{1}{2}\,\Gamma_{\rm D}^2\,\ln(s)$ \\[3pt]
Quantum Fluctuations &\qquad $-u^2\,\ln(s)\T$ &\qquad $\frac{1}{2}\,\Gamma_{\rm D}^2\,\ln(s)$ \\[3pt]
Full One-Loop &\qquad $0\T$ &\qquad $\Gamma_{\rm D}^2\,\ln(s)$ \\[3pt]
\hline\hline
\end{tabular}
\label{table:one}
\end{table}

The interaction correction to the layer pseudospin response function
$\chi_{\rm zz}$ is obtained by closing the scattering function with a
$\sigma^{\rm z}$ vertex at top and bottom.  The $\sigma^{\rm z}$
operator measures the charge difference between $T$ and $B$ layers.
Because it is an effective single-particle theory, fermion
mean-field theory corresponds to response
function diagrams with at most a single particle-hole pair.  It
follows that mean-field theory in BLG is equivalent to a single-loop PRG
calculation in which the BCS and ZS' channels are neglected and only the ZS channel is retained.
In mean-field theory, ideal BLG has
an instability to a state in which charge is spontaneously
transferred between the layers which is signaled by the divergence
of $\chi_{\rm zz}$.  ($\chi_{\rm xx}$ also diverges in mean-field theory but less strongly.)
The PRG analysis demonstrates that the
mean-field theory instability is enhanced by the reinforcing influence of the
ZS' channel contribution.  We discuss pseudospin response functions
which indicate the character of the broken symmetry state further for the physically relevant
$N>1$ case in Section IV.

\begin{widetext}
\subsection{RG Flows for $N=1,2,4$}

Assumming $SU(4)$ invariance, the one-loop flow equations for $N>1$ are derived
in the same way as for the $N=1$ case, except that we need to keep track of three interaction
parameters,
$\Gamma_{\rm S}$ ($\Gamma_{\rm S} = \Gamma_{\rm SDS} = \Gamma_{\rm SSD} = \Gamma_{\rm SDD}$),
$\Gamma_{\rm D}$ ($\Gamma_{\rm D} = \Gamma_{\rm DDS} = \Gamma_{\rm DSD} = \Gamma_{\rm DDD}$)
and
$\Gamma_{\rm X}$ ($\Gamma_{X}=\Gamma_{\rm XDS}=\Gamma_{\rm XSD}=\Gamma_{\rm XDD}$).
A tedious but elementary book-keeping exercise yields the following results:
\begin{eqnarray}
\label{eq:full_rg_flow}
\frac{{\emph d}\,\Gamma_{\rm S}}{\nu_0\,{\emph d}\ln(s)}&=&-\frac{1}{2}\Gamma_{\rm S}^2-(\Gamma_{\rm D}-\Gamma_{\rm X})(\Gamma_{\rm D}-\Gamma_{\rm S})-\frac{N-2}{2}(\Gamma_{\rm D}-\Gamma_{\rm S})^2+\frac{1}{2}(\Gamma_{\rm X}-\Gamma_{\rm S})^2\,,\nonumber\\
\frac{{\emph d}\,\Gamma_{\rm D}}{\nu_0\,{\emph d}\ln(s)}&=&\frac{1}{2}\Gamma_{\rm D}^2+(\Gamma_{\rm D}-\Gamma_{\rm X})(\Gamma_{\rm D}-\Gamma_{\rm S})+\frac{N-2}{2}(\Gamma_{\rm D}-\Gamma_{\rm S})^2-\frac{1}{2}(\Gamma_{\rm D}+\Gamma_{\rm X})^2\,,\nonumber\\
\frac{{\emph d}\,\Gamma_{\rm X}}{\nu_0\,{\emph d}\ln(s)}&=&(\Gamma_{\rm D} - \Gamma_{X}) \Gamma_{\rm X}-\frac{N-2}{2} \Gamma_{\rm X}^2-\frac{1}{2}(\Gamma_{\rm X}-\Gamma_{\rm S})^2-\frac{1}{2}(\Gamma_{\rm X}+\Gamma_{\rm D})^2\,.
\end{eqnarray}
\end{widetext}
This equation is the $SU(4)$ invariant version of the more general ten parameter flow equations
derived in Ref.~\onlinecite{RG_Zhang}. The RG flow equations derived previously in Ref.~\onlinecite{RG_Vafek}
differ only in notation.
The $N=1$ case is recovered by noting that $\Gamma_{\rm S}$ has no physical effect for interactions among
particles of the same flavor and $\Gamma_{\rm D}$ and $\Gamma_{\rm X}$ always enter in the combination
$\Gamma_{\rm D}-\Gamma_{\rm X}$.  Defining $\Gamma_{\rm DS} \equiv \Gamma_{\rm D}-\Gamma_{\rm X}$,
it follows from Eqs.(\ref{eq:full_rg_flow})  that
\begin{eqnarray}
\frac{{\emph d}\,\Gamma_{\rm DS}}{\nu_0\,{\emph
d}\ln(s)}&=&\Gamma_{\rm DS}^2+\frac{N-1}{2} \, (\Gamma_{\rm
D}-\Gamma_{\rm S})^2+\frac{N-1}{2} \, \Gamma_{\rm X}^2\, ,\nonumber\\
\end{eqnarray}
yielding a one-parameter flow equation for the $N=1$ case. Note that this $\Gamma_{\rm DS}$ also represents $\Gamma_{\rm DSS}$ in the $N=4$ case.\cite{RG_Zhang}

The only fixed point of these RG flow equations is the non-interacting
one.  The character of the broken symmetry state can be estimated by following the
RG flows to strong interactions, or more reliably by following the flow equations for
appropriate susceptibilities as discussed in the next section.   In order to represent the property
that bare same layer interactions should be slightly stronger than the bare
different layer interactions and that the bare layer interchange interaction
vanishes, we start the RG flow integrations from
\begin{eqnarray}
\label{eq:bare}
V_{\rm S}> V_{\rm D}, \quad V_{\rm X}=0\,.
\end{eqnarray}
When the flows start from an interaction with these properties,
the coupling constants diverge simultaneously at a finite value of $\ln(s)$, and for $N=2,4$
satisfy the following properties at the
divergence point:
\begin{eqnarray}
\label{eq:rg_results}
&&\frac{\Gamma_{\rm S}}{\Gamma_{\rm D}} = -1,\nonumber\\
&&\frac{\Gamma_{\rm S}}{\Gamma_{\rm X}}\simeq -0.5425\;(\text{for}\; N=2)\,,\nonumber\\
&&\frac{\Gamma_{\rm S}}{\Gamma_{\rm X}}\simeq -0.2624\;(\text{for}\; N=4)\,.
\end{eqnarray}
The value of the first ratio in Eq.(\ref{eq:rg_results}) follows from the RG flow equation for $\Gamma_{\rm S}+\Gamma_{\rm D}$ which satisfies
\begin{equation}
\frac{{\emph d}\,(\Gamma_{\rm S}+\Gamma_{\rm D})}{\nu_0\,{\emph d}\ln(s)}=-\Gamma_{\rm X}(\Gamma_{\rm S}+\Gamma_{\rm D}).
\end{equation}
The rate of growth of $\Gamma_{\rm S}+\Gamma_{\rm D}$ is proportional to $|\Gamma_{\rm X}|$, whereas
$\Gamma_{\rm S}$ and $|\Gamma_{\rm X}|$ grow like $|\Gamma_{\rm X}|^2$.
Equivalent results in a different notation, using $2g_{\rm 0,z}=\Gamma_{\rm S}\pm\Gamma_{\rm D}$ and $2g_{\perp}=\Gamma_{\rm X}$, have been derived previously by Vafek and Yang.\cite{RG_Vafek}
These results do not depend on the bare coupling constants
as long as the physical conditions specified by Eq.(\ref{eq:bare}) are satisfied.
Note that the RG flows for $N=4$ are still qualitatively different from those of the $N \to \infty$
limit.

\begin{figure}[htbp]
\centering \scalebox{0.8} {\includegraphics*[2.2in,2.3in][6.0in,8.8in]{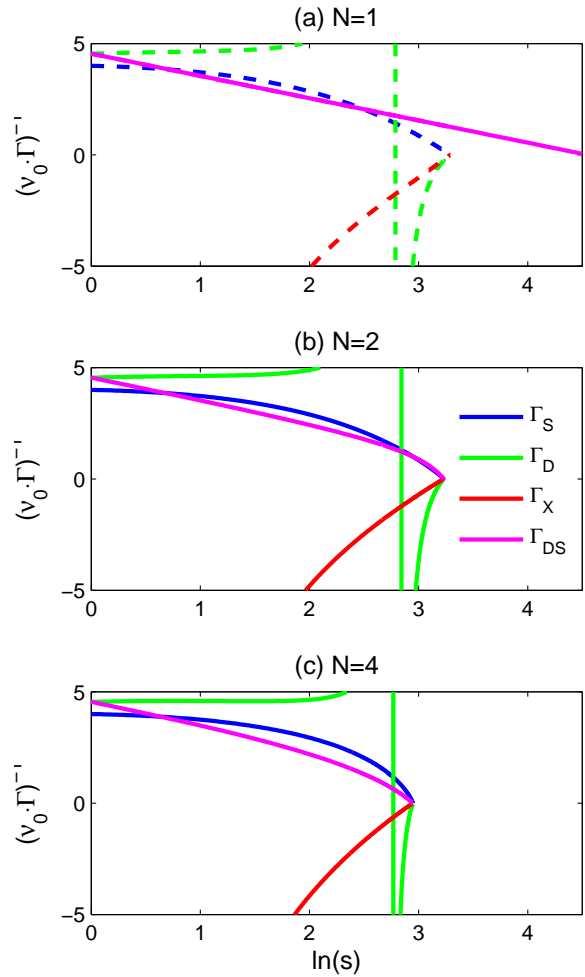}}
\caption{\label{fig:rgflow}(Color online) RG flows of the interaction couplings for (a) $N=1$, (b) $N=2$ and (c) $N=4$. In the $N=1$ case the dashed lines are unphysical since interactions always enter observables
in the combination $\Gamma_{\rm DS}=\Gamma_{\rm D}-\Gamma_{\rm X}$.}
\end{figure}

Typical RG flows, obtained by integrating Eq.(\ref{eq:full_rg_flow}) numerically, are plotted in Fig.~\ref{fig:rgflow}.
We note that, when combined with the mean-field-theory of the gapped state,\cite{SQH_Zhang,prepare} the value
$\nu_0V_{\rm S}=0.25$ yields values of the spontaneous gap, the critical temperature, and the
carrier density at which order is absent that are consistent with some recent experimental observations\cite{AB_Lau,AB_Lau1}.
We therefore choose to use the bare interaction parameters
$\nu_0V_{\rm S}=0.25$ and $\nu_0V_{\rm D}=0.22$ to illustrate typical RG flows.
We find that the interaction parameters flow away from the non-interacting fixed point and
diverge at a finite value of $s$.
The instability criterion implied by the one-loop RG calculation is $\nu_0 V_{\rm S} \simeq 0.7 / \ln(s)$ for $N=4$,
{\rm i..e} the instability tendency is enhanced compared to the $N=1$ case.
(Although the $\Gamma_{\rm S}$,
$\Gamma_{\rm D}$ and $\Gamma_{\rm X}$ interaction parameters do diverge at a similar value of
$\ln(s)$ in the $N=1$ case, they have no physical effect because they enter observables
in the combination $\Gamma_{\rm DS}=\Gamma_{\rm D}-\Gamma_{\rm X}$.)
Because of the quadratic band dispersion,
the ratio of the interlayer coupling scale to the gap ($\gamma_{1}/\Delta$)
can be associated \cite{LAF_Zhang} with the square of the momentum scaling factor
$s$ at the phase transition.  This yields a value of $\Delta$ on the order of $1$ meV,
close to the scale of the gap values seen in some experiments.\cite{Yacoby1,Yacoby2,AB_Schonenberger,AB_Lau,AB_Lau1}

\section{Susceptibilities}

More insight into the likely nature of the broken symmetry state which occurs in BLG can be
obtained by using the PRG calculations to estimate the long wavelength static limit of the pseudospin
susceptibilities:
\begin{equation}
\chi_{\rm T\gamma',B\gamma}({\bm{q}},i\omega)=\int_{0}^{\beta} d\tau e^{\rm i\omega\tau}\left<T_{\tau}S_{\rm T\gamma'}({\bm{q}},\tau)S_{\rm B\gamma}({-\bm{q}},0)\right>\,,
\end{equation}
where $T,B=x,y,z$ label layer-pseudospin components (instead of top and bottom layers) and $\gamma',\gamma$ label either a unit matrix ($\gamma=0$)
or a generator of rotations ($\gamma=1, \ldots, 15$) in the $4$-component valley-spin flavor space:
\begin{equation}
S_{\rm T\gamma}({\bm {q}},\tau)= \sum_{{\bm {k}}}c_{{\bm {k+q}},{\rm \alpha i'}}^{\dagger} \sigma^{\rm (T)}_{\rm \alpha\beta}
A^{\rm \gamma}_{\rm i'i} c_{{\bm {k}},{\rm \beta i}}\,.
\label{eq:chi}
\end{equation}
\begin{figure}[t]
{\scalebox{0.7}{\includegraphics*[1.9in,7.25in][6.6in,10.3in]{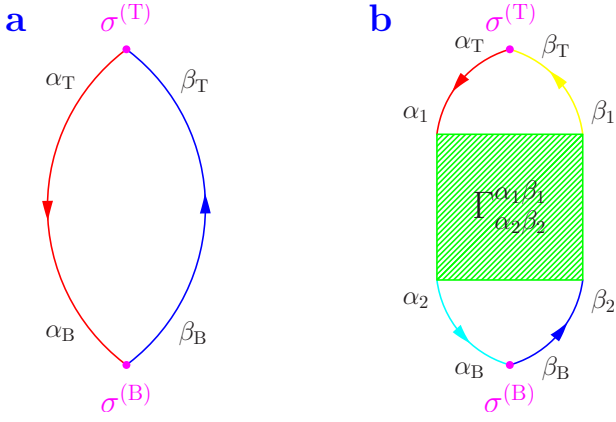}}}
\caption{\label{fig:susc}(Color online) Feynman diagrams for the pseudospin susceptibilities. {\bf (a)} The non-interacting susceptibility and {\bf (b)} the interaction correction to the susceptibility. $T,B=x,y,z$ label layer-pseudospin components.
The spin-valley flavor labels are not explicitly indicated.}
\end{figure}
A physically transparent choice\cite{su4} for the $SU(4)$ generators is to set $A^{\rm \gamma}_{i',i}$ to a
Pauli matrix which acts on the spin degree-of-freedom for $\gamma=1,2,3$, to a Pauli matrix which
acts on the valley degree-of-freedom for $\gamma=4,5,6$, and to a product of spin and valley Pauli
matrices for $\gamma=7,\ldots,15$.  It follows from $SU(4)$ invariance\cite{su4} that the normal state
susceptibilities in this case are $ \propto \delta_{\rm \gamma',\gamma}$ and equal for all values of $\gamma \ge 1$.
Below we refer to the $\gamma=0$ susceptibility ($\chi^c$) as the flavor charge susceptibility and to the
$\gamma \ne 0$ susceptibility ($\chi^s$) as the flavor spin susceptibility.
The conservation of particle-number in
each layer at long wavelengths, explained previously, implies that $\chi_{\rm T\gamma,B\gamma} \propto \delta_{\rm T,B}$
and that $\chi_{\rm x\gamma,x\gamma} = \chi_{\rm y\gamma,\rm y\gamma}$.
Divergences in $\chi^{\rm c}_{\rm z,z}$ signal ordered states in which the charge density is transferred between layers
in the same sense for all flavors, while divergences in $\chi^{\rm s}_{\rm z,\rm z}$ signal flavor dependent layer-inversion
symmetry breaking with no overall charge transfer.  Divergences in $\chi^{\rm c,s}_{\rm x,\rm x}$ for $T=x,y$, on the other
hand imply broken symmetry states in which the phase relationship between layers is altered and rotational
symmetry within the layers is broken.  Broken layer-inversion symmetry states have an energy gap for charged excitations
and large momentum space Berry curvatures that lead to large anomalous Hall effect contributions
from individual flavors.\cite{SQH_Zhang}  These states will be referred to below as spontaneous quantum Hall states.\cite{SQH_Zhang}
The broken rotational symmetry state will be referred to below as the nematic state.

Provided that the system has a single continuous phase transition, the ordered state
character can in principle be determined by identifying the pseudospin susceptibility which diverges first
as temperature is lowered.
Although we cannot make that determination definitively from the RG interaction strength flows,
we can obtain some guidance by examining how the interaction contributions to the susceptibilities
grow under the RG flows.
The pseudospin response functions of interest can be related to the renormalized interactions as summarized in
the Feynman diagrams of Fig.~\ref{fig:susc}.
Fig.~\ref{fig:susc}(a) represents the non-interacting pseudospin susceptibility ${\chi}^{0}$ and Fig.~\ref{fig:susc}(b) represents the interaction correction ${\chi}^{\rm int}$. In the long wavelength static limit the
explicit expressions for these susceptibilities are:
\begin{widetext}
\begin{eqnarray}
\!\!\!\!\!\!\!\!\!\chi_{\rm TB}^{0}&=&-\!\!\!\int\!\!\! \frac{d^2 {\bm q}}{(2\pi)^2 (\beta\hbar^2)}\sum_{\omega}
\sum_{\alpha_{\rm T}\beta_{\rm T}}\sum_{\alpha_{\rm B}\beta_{\rm B}}
{\mathcal G}_{\alpha_{\rm B}\alpha_{\rm T}}({\bm{q}},i\omega)\sigma^{\rm (T)}_{\alpha_{\rm T}\beta_{\rm T}}
{\mathcal G}_{\beta_{\rm T}\beta_{\rm B}}({\bm{q}},i\omega)\sigma^{\rm (B)}_{\beta_{\rm B}\alpha_{\rm B}}\,,\\
\label{eq:chi0}
\!\!\!\!\!\!\!\!\!\chi_{\rm TB}^{\rm int}&=&-\!\!\!\int\!\!\!\frac{d^2 {\bm q_1} d^2 {\bm q_2}}{(2\pi)^4(\beta\hbar^2)^2}\sum_{\omega_{1}\omega_{2}}
\sum_{\rm all\,\alpha\beta}
{\mathcal G}_{\alpha_1\alpha_{\rm T}}({\bm{q_1}},i\omega_1)\sigma^{\rm (T)}_{\alpha_{\rm T}\beta_{\rm T}}{\mathcal G}_{\beta_{\rm T}\beta_1}({\bm{q_1}},i\omega_1)
\Gamma^{\beta_1\alpha_1}_{\alpha_2\beta_2}
{\mathcal G}_{\beta_2\beta_{\rm B}}({\bm{q_2}},i\omega_2)\sigma^{\rm (B)}_{\beta_{\rm B}\alpha_{\rm B}}{\mathcal G}_{\alpha_{\rm B}\alpha_2}({\bm{q_2}},i\omega_2)\,.
\label{eq:chiI}
\end{eqnarray}
\end{widetext}
Note that in these equations all propagators are diagonal in the implicit spin-valley flavor labels.

In the noninteracting case, the inter-flavor (D) susceptibilities vanish because the single-particle Hamiltonian is
diagonal in flavor and the intra-flavor (S) susceptibilities are positive because of the band state pseudospin structure:
\begin{eqnarray}
\chi_{\rm zz}^{\rm 0D}&=&\chi_{\rm xx}^{\rm 0D}=\chi_{\rm yy}^{\rm 0D}=0\,,\nonumber\\
\chi_{\rm zz}^{\rm 0S}&=&2\chi_{\rm xx}^{\rm 0S}=2\chi_{\rm yy}^{\rm 0S}= 2\nu_0\ln(s) .
\end{eqnarray}
These pseudospin susceptibilities capture the contribution of momentum $\bm{q}=0$ vertical
interband quantum fluctuations involving states with
energies measured from the Dirac point between $\gamma_1$ and $\gamma_1/s^2$.
The factor of two difference between $\chi_{\rm zz}^{\rm 0S}$ and $\chi_{\rm xx}^{\rm 0S}$ demonstrates that
the band state is more susceptible to a gap opening perturbation than to a nematic perturbation.
This property can be understood \cite{MF_MacDonald} in terms of the pseudospin orientations of the band states; in
particular a $\hat{z}$ pseudospin effective field is perpendicular to the valence band pseudospin for
all momentum orientations, so that $\hat{z}$ direction fields always yield a strong response.
On the other hand $\hat{x}-\hat{y}$-plane pseudospin fields are not in general perpendicular to the
valence band pseudospin and consequently produce a weaker response averaged over momentum
orientations.  The larger response to a $\hat{z}$ pseudospin effective field is related to the
well-known property of BLG that a potential difference between the top and bottom layers lead to an energy gap at the Dirac point.  The corresponding charge susceptibility (density-density channel), obtained by replacing the pseudospin-field
Pauli matrix by an identify matrix, vanishes
because conduction and valence band states are orthogonal.

When the interactions flow to strong values, the susceptibilities are dominated by their interaction
contributions. When these are included we find that
\begin{eqnarray}
\label{eq:inter}
\chi_{\rm zz}^{\rm D}&=&2(\nu_0\ln(s))^2(\Gamma_{\rm D}-\Gamma_{\rm S})\,,\nonumber\\
\chi_{\rm xx}^{\rm D}&=&-\frac{1}{2}(\nu_0\ln(s))^2\Gamma_{\rm X}\,,\\
\nonumber\\
\label{eq:intra}
\chi_{\rm zz}^{\rm S}&=&2\nu_0\ln(s)+2(\nu_0\ln(s))^2 \, (\Gamma_{\rm D} - \Gamma_{X}) \,,\nonumber\\
\chi_{\rm xx}^{\rm S}&=&\nu_0\ln(s)+\frac{1}{2}(\nu_0\ln(s))^2 (\Gamma_{\rm D} - \Gamma_{X} )\,.
\end{eqnarray}
Note that the large susceptibility in $\chi_{\rm xx}^{\rm D}$ is attribute to the layer interchange processes $\Gamma_{\rm X}$ while the strong divergences in $\chi_{\rm zz,xx}^{\rm S}$ are due to intra-flavor exchange interactions, recalling $\Gamma_{\rm D}-\Gamma_{\rm X}\equiv\Gamma_{\rm DS}$ in Eq.(\ref{eq:intra}).

For the $N=1$ case there is no different-flavor response and we therefore find that near the instability
\begin{eqnarray}
\chi_{\rm zz} = 4 \chi_{\rm xx}\, =  4 \chi_{\rm yy}\,,
\end{eqnarray}
suggesting that a gapped spontaneous quantum Hall state is most likely.

For the physical $N=4$ case we can relate the flavor charge and flavor spin response functions to $\chi^{\rm S}$ and
$\chi^{\rm D}$ by examining the particular case of responses to perturbations that are diagonal in flavor index,
{\em i.e.} perturbations proportional to $s^{\rm z}$, $\tau^{\rm z}$ or $s^{\rm z} \tau^{\rm z}$.
We find that near the divergence point
\begin{eqnarray}
\chi^{\rm c}_{\rm zz} &=& \frac{1}{4} \, \chi_{\rm zz}^{\rm S} + \frac{3}{4} \, \chi_{\rm zz}^{\rm D}\,,\\
\chi^{\rm s}_{\rm zz} &=& \frac{1}{4} \, \chi_{\rm zz}^{\rm S} - \frac{1}{4} \, \chi_{\rm zz}^{\rm D}\,.
\end{eqnarray}
$\chi^{\rm c,s}_{\rm xx}$ have the same relations to $\chi^{\rm S}_{\rm xx}$ and $\chi^{\rm D}_{\rm xx}$.
Combining Eq.(\ref{eq:inter}) and Eq.(\ref{eq:intra}) with the results from  Eq.(\ref{eq:rg_results}) for $N=4$
we find that
\begin{eqnarray}
\label{eq:N4p}
\chi^{\rm c}_{\rm xx} &\to&  +1.7805 \,\, \Gamma_{\rm S} (\nu_0 \ln(s))^2\,, \nonumber  \\
\chi^{\rm s}_{\rm zz} &\to&  +2.4055 \,\, \Gamma_{\rm S} (\nu_0 \ln(s))^2\,,  \\
\nonumber\\
\label{eq:N4n}
\chi^{\rm c}_{\rm zz} &\to&  -1.5945 \,\, \Gamma_{\rm S} (\nu_0 \ln(s))^2\,, \nonumber \\
\chi^{\rm s}_{\rm xx} &\to&  -0.1250 \,\, \Gamma_{\rm S} (\nu_0 \ln(s))^2\,.
\end{eqnarray}
The property that all susceptibilities diverge together is an artifact of the single-loop RG calculation
which becomes questionable when the scaled interactions are strong.  Still, the fact that the $\chi^{\rm s}_{\rm zz}$
diverges most strongly suggests that the spin-channel gapped state is more likely than the
charge channel nematic state.
(Negative coefficients in Eq.(\ref{eq:N4n})  imply that
one-loop interaction corrections tend to reduce the susceptibilities
$\chi_{\rm zz}^{\rm c}$ and $\chi_{\rm xx}^{\rm s}$ rather than to increase them and
therefore do not suggest instabilities.)

\begin{figure}[t]
\centering \scalebox{0.72} {\includegraphics*{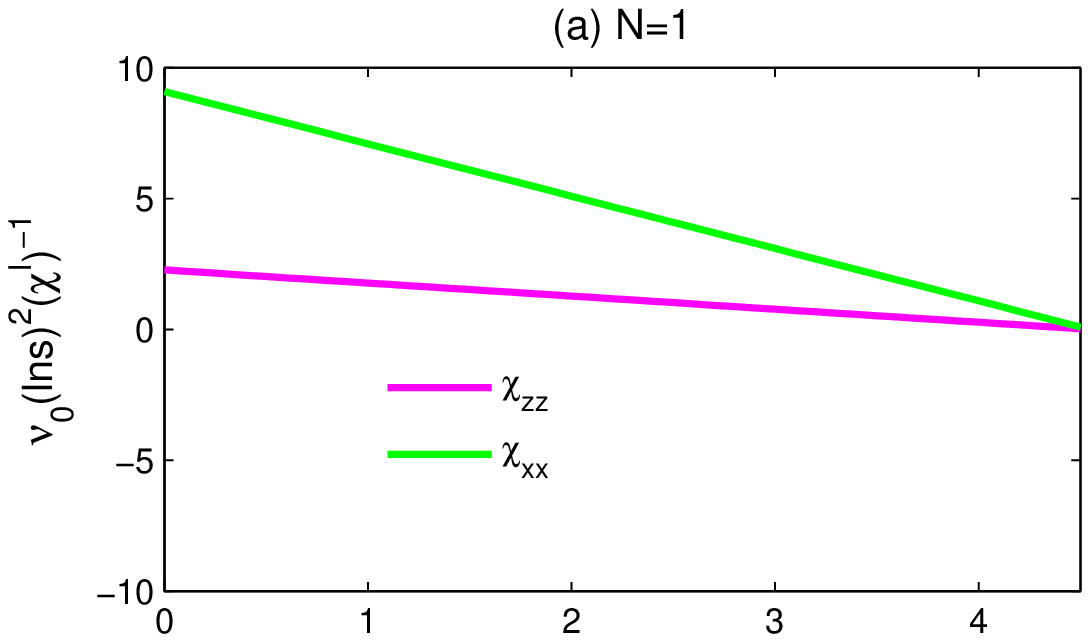}}
\scalebox{0.72} {\includegraphics*{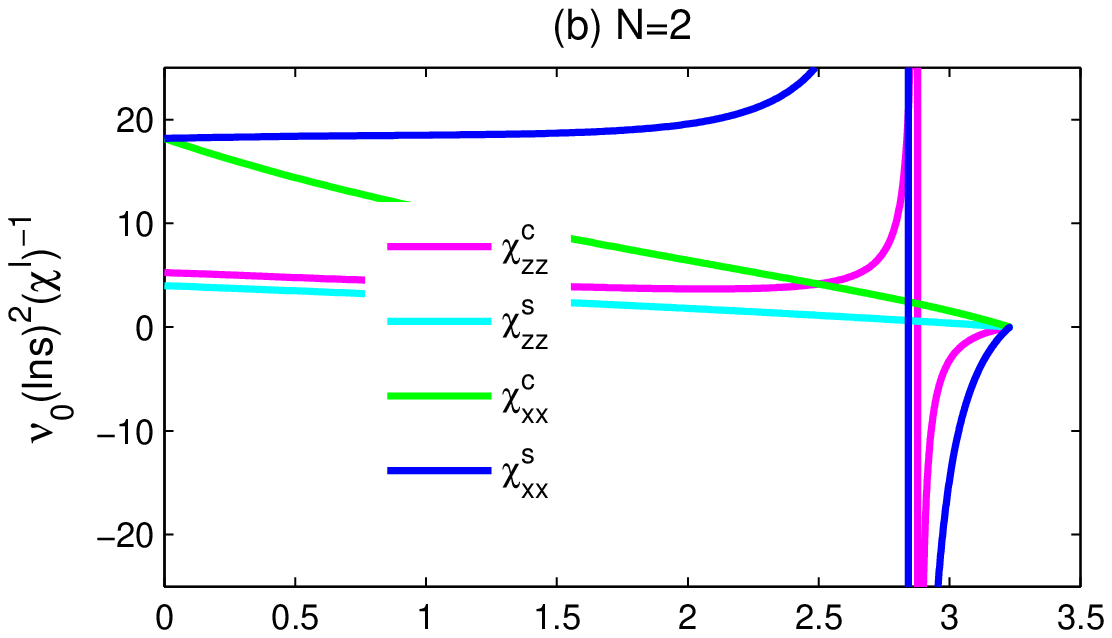}}
\scalebox{0.72} {\includegraphics*{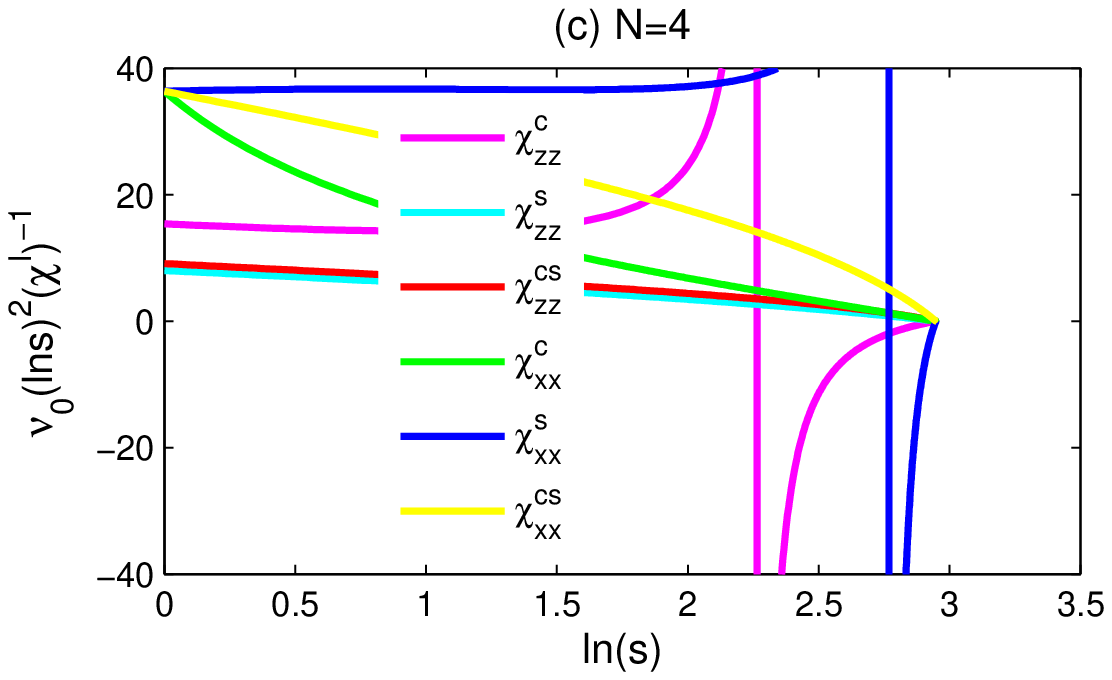}}
\caption{\label{fig:susflow}(Color online) RG flows of interacting susceptibilities for different ordering tendencies. (a) $N=1$, (b) $N=2$ and (c) $N=4$. We use the dimensionless quantity $\nu_0(\ln(s))^2(\chi^{\rm T})^{-1}$ in the figures.}
\end{figure}

In the $N=2$ case, which accounts for only spin or valley but not both, $\chi_{\rm T\gamma,B\gamma'}$
the $4$ dimensional $\gamma$-matrices are replaced by $2$-dimensional Pauli matrices. We find that
\begin{eqnarray}
\chi_{\rm zz}^{\rm c}&=&\frac{1}{2}\,\chi_{\rm zz}^{\rm S}+\frac{1}{2}\,\chi_{\rm zz}^{\rm D}\,,\\
\chi_{\rm zz}^{\rm s}&=&\frac{1}{2}\,\chi_{\rm zz}^{\rm S}-\frac{1}{2}\,\chi_{\rm zz}^{\rm D}\,,
\end{eqnarray}
which equally applies to the $\chi_{\rm xx}$ channel. Combining the $N=2$ results in Eq.(\ref{eq:rg_results})
with Eq.(\ref{eq:inter}) and Eq.(\ref{eq:intra}) we obtain
\begin{eqnarray}
\chi^{\rm c}_{\rm zz} &\to&  -1.1567 \,\, \Gamma_{\rm S} (\nu_0 \ln(s))^2\,, \nonumber \\
\chi^{\rm s}_{\rm zz} &\to&  +2.8433 \,\, \Gamma_{\rm S} (\nu_0 \ln(s))^2\,, \nonumber \\
\chi^{\rm c}_{\rm xx} &\to&  +0.6717 \,\, \Gamma_{\rm S} (\nu_0 \ln(s))^2\,, \nonumber  \\
\chi^{\rm s}_{\rm xx} &\to&  -0.2500 \,\, \Gamma_{\rm S} (\nu_0 \ln(s))^2\,.
\end{eqnarray}

In both $N=2,4$ cases interactions favor breaking layer-inversion symmetry in the flavor spin channel
and orientational symmetry in the flavor charge channel.
The strongest divergence occurs for the flavor spin channel broken layer-inversion symmetry state.\cite{RG_Zhang}
The $N=1$ susceptibility and the flavor charge and spin channel susceptibilities for $N=2,4$ are
illustrated in Fig.\ref{fig:susflow} and agree with the above analytic analysis.
Because $\chi_{\rm zz}^{\rm s}$ is always the most {\em positvely} divergent channel
we conclude that the dominant many-body effect in BLG is spin-channel spontaneous layer-inversion
symmetry breaking yielding BLG states with spontaneous charge gaps.
These results verify our original theoretical predictions \cite{MF_Min,RG_Zhang} and are consistent with
recent experimental observations in dual-gated ultra-clean suspended BLG devices.\cite{Yacoby1,Yacoby2,AB_Lau,AB_Lau1}

Finally, we point out that an interlayer potential breaks $SU(4)$ invariance in the $N=4$ case, leading to a range of stability for a state in which three flavors are polarized toward one layer and one flavor toward the other layer when an ordered state appears.  For instance, a possible mass term in a mean-field theory with broken layer-inversion symmetry can be proportional to $(1+s^{\rm z}+\tau^{\rm z}-s^{\rm z}\tau^{\rm z})\,\sigma^{\rm z}/2$. This gapped state breaks parity inversion and time reversal symmetries, favored by small interlayer electric and magnetic fields \cite{SQH_Zhang,AB_Lau} that break $SU(4)$ symmetry. The difference between this state and other spontaneously broken symmetry states can be further appreciated as follows in our full model with {\em approximated} $SU(4)$ symmetry. This state reduces the full symmetry to $SU(3)\times SU(1)\times U(1)$ with three Goldstone modes, in contrast to the three possible states in flavor spin channel that break the symmetry down to $SU(2)\times SU(2)\times U(1)$ with four Goldstone modes, and there is no soft mode in the flavor charge channel.

\section{Discussion}

In this paper we have attempted to shed light on the pseudospin ferromagnet
ground state of BLG by carrying out a conventional perturbative renormalization group
calculation for a model with $SU(4)$ invariance in spin-valley space, and examining interaction corrections to
spin and charge pseudospin susceptibilities.  The PRG flows lead to strong layer interchange
scattering processes which favor gapless nematic ($\hat{x}-\hat{y}$-plane pseudospin ferromagnet)
ground states, competing with the gapped ($\hat{z}$-direction pseudospin ferromagnet)
states favored by intra-flavor exchange interactions and predicted \cite{MF_Min} by mean field theory.
The role of exchange is especially dominant in the single-flavor $N=1$ case in which
the PRG flow equations lead to divergent interactions which vanish when projected onto the many-fermion
Hilbert space.  For $N=1$ the interactions enter observable only in the
combination $\Gamma_{\rm DS}=\Gamma_{\rm D}-\Gamma_{\rm X}$
that is uniquely allowed by the Pauli exclusion principle.

We conclude that the most likely ground states
are gapped spin-channel Ising pseudospin ferromagnets, in agreement with
some \cite{AB_Schonenberger,AB_Lau,AB_Lau1} recent experiments, although the competition
becomes closer when the number of flavors is larger, and quite close for the physically
relevant $N=4$ case.  According to our estimates, broken symmetry states appear
at energy scales slightly larger than those at which trigonal warping effects in the
single-particle Hamiltonian \cite{AB_McCann} start to have a large influence on the
band structure, splitting the vorticity $J=2$ vortex of the chiral model into
four $J=1$ Dirac points.  The strength of
our conclusions is limited by our neglect of the long-range character of Coulomb interactions
and by the fact that our one-loop PRG calculations become unreliable once the
effective interactions flow to strong values, {\em i.e.} once $\nu_0 \Gamma \gtrsim 1$.

The details of our calculations are conventional and in particular evaluate the
loop corrections to scattering amplitudes for the case in which both incoming and
out-going momenta are at the Dirac point.  This simplification is normally justified by the
irrelevance of gradient corrections to the effective interactions.  As we now explain, in the
case of BLG, this approximation likely overestimates the strength of the $\Gamma_{X}$
layer interchange interactions.  Consider for example the ZS process which is non-zero because
of the triplet layer off-diagonal contribution to the Green's functions, as we have approximated
in Eq.(\ref{eq:ZS}).  When the incoming electrons have finite momenta $\bm{k}$ and $\bm{p}$, the
momentum arguments of the virtual states in the loop diagram are $\bm{k+q}$ and $\bm{p+q}$.
The triplet propogator phase factors which are integrated over $\bm{q}$ in the loop correction to
$\Gamma_{X}$ therefore becomes $e^{\pm 2i\theta}$ where $\theta = \phi_{\bm{k+q}}-\phi_{\bm{p+q}}$.
Notice that this phase factor has a singular dependence on $\bm{k}$ and $\bm{p}$ at
small $\bm{q}$ and that the approximation we have used previously, in which we replace
$\theta$ by zero, is valid only for $q \gg p,k$.  When the phase factors are taken into account
the $\Gamma_{X}$ scattering amplitude has an additional phase factor which will average
to a value smaller than one in physical observables like the pseudospin susceptibilities.
It therefore appears to us that our calculation likely overestimates the strength of
the interactions which favor the nematic state.

The gapped states have the interesting property that they support large
momentum space Berry curvatures,\cite{Thouless,SQH_Zhang} which can lead to interaction-induced
quantum Hall effects.  The regions of momentum space near
both K and K' valleys flavor contribute $\sim \pm e^2/h$ to the
Hall conductivity for each spin.  When all flavors make a contribution of the
same sign, the total Hall conductivity is exactly $\pm 4 e^2/h$.
This gapped bilayer state is therefore a quantized anomalous Hall (QAH)
insulator \cite{SQH_Zhang,SQH_Levitov,Thouless,AH_Haldane} with chiral edge states.
Because of its nontrivial overall orbital moment\cite{SQH_Zhang} the QAH state is favored by a perpendicular magnetic field. The state in which the sense of layer polarization is opposite for
opposite spins and for opposite valleys does not break time-reversal symmetry
and has a mean-field interaction terms with an effective spin-orbit coupling \cite{QSH_Kane}
that leads to a quantum spin Hall (QSH) state with zero charge Hall conductivity.
The QSH state has helical edge states protected by time-reversal symmetry.
Its $\mathcal{Z}_2$ classification in an interacting $N$-layer chiral graphene is given by\cite{SQH_Zhang} $\nu=N\mod 2$.
When lattice effects \cite{MF_Jung} are taken into account, inter-valley exchange
weakly favors the layer-antiferromagnet (LAF) \cite{SQH_Zhang} state.
(That observation does not definitively make the case for this particular state, since
inter-valley exchange could be less important than correlation effects.)
For a LAF state, $\sigma_{\rm H}=0$ even though time-reversal symmetry is broken
by opposite spin polarizations on the top and bottom layers.
In a continuum model that neglects lattice effects and has $SU(4)$ symmetry, the
three states are all members of the continuous family of degenerate broken symmetry states
that is signaled by divergence of $\chi_{\rm zz}^{\rm s}$.

In addition to their difference in edge state properties,
the QAH, QSH and LAF states also exhibit different responses \cite{LAF_Zhang} to Zeeman fields
which couple to spin and can be realized in the absence of
orbital coupling by applying magnetic fields parallel to graphene layers.
In the QAH case, the ground state is unchanged but the quasiparticle gap is
reduced, vanishing when the Zeeman-coupling strength is equal to the ground state gap via a mechanism reminiscent of the Clogston limit in superconductors.
The QSH and LAF states respond to Zeeman fields by establishing noncollinear spin states within each valley
and evolving toward an unusual kind of exciton condensate in the strong Zeeman-coupling limit.
The quasiparticle gap of QSH and LAF states is however independent of Zeeman-coupling strength
drawing a sharp distinction with the QAH case.
It is therefore possible to use Zeeman responses and edge state signatures to identify the character of the bilayer ground state experimentally.\cite{LAF_Zhang}
On the other hand, the three states respond very similarly to an electric field between the layers, which can induce first order transitions at which the total layer polarization jumps.\cite{AB_Lau1,LAF_Zhang}
This state favored by an interlayer electric field is a quantum valley Hall state with zero Hall conductivity but, in the absence of scattering which mixes valleys counterpropagating valley-resolved edge states.
We note that a small electric field between the layers can also possibly stabilize
 a state in which one flavor is polarized in a sense
opposite to the other three and charge, valley, and spin Hall conductivities are all nonzero.\cite{SQH_Zhang}

One consequence of the above mentioned spontaneous quantum Hall effects \cite{SQH_Zhang} in each valley is that the gapped states have a
very simple adiabatic evolution with magnetic field $B$ in which the gap
remains open, but the carrier density at which the gap appears varies with
field.  The total filling factor inside the gap at finite $B$  is $\nu$ for a state with a
Hall conductance equal to $\nu e^2/h$ at zero field.  The topological
character \cite{SQH_Zhang} of the gapped states in BLG is induced \cite{RG_Zhang} by weak bare electron-electron interactions and is enhanced by the nearly flat bands with chiral pseudospin textures, and is distinct from the $\mathcal{Z}_2$ topological insulators \cite{QSH_Kane,Kane_RMP,Zhang_RMP} that require strong intrinsic spin-orbital coupling.
Although interaction-induced topological states apparently do not
occur in single-layer graphene, as proposed \cite{QSH_Raghu} on the basis of
extended Hubbard models with at least finite interaction strength, it appears
quite possible that they do occur in bilayers via a weak interaction instability.\cite{RG_Zhang}

There are three previous articles \cite{RG_Zhang,RG_Vafek,RG_Falko} which discuss similar RG calculations.
Ref.~\onlinecite{RG_Vafek} reports equivalent RG flow equations for the $SU(4)$ invariant model in terms of
interaction parameters that are related to ours by $g_{\rm 0,z}=\Gamma_{\rm S}\pm \Gamma_{\rm D}$ and $2g_{\perp}=\Gamma_{\rm X}$ .
Several references have started their RG flows from effective interactions which include symmetry-allowed corrections due to lattice effects.
Ref.~\onlinecite{RG_Falko} attempts to account for the long-range of the Coulomb interaction by explicitly accounting for screening
and then replacing the Coulomb amplitude by an average value, and estimates the ground state by applying mean-field theory to interactions
from PRG flows that have been truncated at an intermediate interaction strength.
All calculations recognize the competition between nematic and gapped states, but small differences in approximations or in estimates of
bare coupling parameters have lead to different final conclusions.
We note that the recent elaborations\cite{Vafek_2,Falko_2}
of the calculations which motivated nematic state proposals\cite{RG_Vafek,RG_Falko}
recognize that gapped states\cite{RG_Zhang} are also a possibility.

Although these calculations, like ours, shed light on the physics at play in determining the ground state,
it is difficult to make definitive conclusions based on theory alone and we must ultimately appeal to experiment.
Here the situation is also confusing.  Some experimental studies \cite{AB_Schonenberger,AB_Lau,AB_Lau1}
of high-quality suspended bilayers appear to reveal gaps whose size, carrier-density-dependence, and temperature-dependence, is roughly consistent with mean-field-theory predictions.\cite{AB_Lau1,prepare}
These observations are consistent with the
conclusions of this paper.  On the other hand this behavior is not observed in all samples, even in
samples which appear to be similar to the gapped samples.  One experimental study\cite{Geim}
focuses on the magnetic-field $B$ dependence of the gap at filling factor $\nu=4$ in a sample which
does not show a gap at $B=0$.  The anomalous weak $B$ persistence of the $\nu=4$ gap in this experiment
appears to be consistent with earlier studies by Yacoby and collaborators.\cite{Yacoby1,Yacoby2}
Although interpreted as evidence for the nematic study, the persistence of the $\nu=4$
gap is, it appears to us, equally consistent with a field-driven crossover or transition to a
gapped $\sigma_{\rm H} = 4 e^2/h$ state.  Although this gap is likely not the ground state at $B=0$ it
is favored at filling factor $\nu=4$, because its gap is then pinned to the Fermi level.
In this interpretation the increased conductance observed at weak fields
in Ref.~\onlinecite{Geim} would be associated with a first order
transition \cite{AB_Lau1,Dagim} between $\sigma_{\rm H} = 4 e^2/h$ and
$\sigma_{\rm H} =0$ states which induces a network of current-supporting \cite{Qiao} domain walls.\cite{SQH1_Zhang}

Similar spontaneously broken symmetry physics\cite{RG_Zhang} is also likely in
thicker high quality graphene films with
chiral (ABC) stacking order.\cite{SQH_Zhang,ABC_Lau}
In an $N$-layer chiral stack,\cite{ABC_Zhang} the low-energy sublattice sites are localized in the
outermost layers, $A_1$ and $B_{\rm N}$.
Hopping occurs between these two sites via an $N$-step virtual process
which leads to $p^{\rm N}$ energy dispersions and an $N\pi$ Berry phase.
For larger $N$, the low-energy bands are increasingly flat and the pseudospin chirality
larger, at least when weak remote hopping processes are neglected, leading to
larger opening for many-body interaction effects.
In the simplified chiral model, the density of states $D(E)\sim E^{\rm (2-N)/N}$ diverges as $E$ approaches zero for $N>2$ whereas it remains finite for $N=2$. In the PRG language, this difference
corresponds to the fact that the short-range interactions at tree level are marginal for $N=2$ but relevant
for $N>2$.  Since
$V_{\rm X}=0$ at tree level, as discussed in Section III-C,
it seems that gapped spontaneous quantum Hall states \cite{SQH_Zhang}
which break layer-inversion symmetry and produce large momentum space Berry curvatures can
occur in multilayers as well.  Future experimental and theoretical work will
be necessary to sort out the competition between interactions and remote hopping terms
in the Hamiltonian.

\section{Acknowledgements}
FZ has been supported by DARPA under grant SPAWAR N66001-11-1-4110.
AHM acknowledges valuable discussions with Joel Moore on the charge and spin susceptibilities of
$SU(4)$ invariant systems.
AHM was supported by Welch Foundation grant TBF1473, NRI-SWAN, and DOE Division of
Materials Sciences and Engineering grant DE-FG03-02ER45958.


\begin{thebibliography}{100}
\bibitem{ft_inversion}
 We refer here to the valley-projected layer-inversion symmetry. This symmetry is broken in the sense that terms proportional to $\sigma^{\rm z}$ are generated by electron-electron interactions in the mean-field continuum theory. Note that this layer-inversion symmetry is different than the parity inversion symmetry with the operator $\mathcal{P}=\tau^{\rm x}\sigma^{\rm x}$. The band Hamiltoian of chiral graphene has parity inversion symmetry, as $\mathcal{P}\mathcal{H}_{\rm J}({\bm q})\mathcal{P}^{-1}=\mathcal{H}_{\rm J}(-{\bm q})$. All the spontaneously gapped states classified in ref.~\onlinecite{SQH_Zhang} break the valley projected layer-inversion symmetry. However, the quantum anomalous Hall and quantum spin Hall states preserve the parity inversion symmetry.
\bibitem{ft_valley}  Specifically we rotate by $180^{\circ}$ around the $\hat{y}$-axis so that
$\sigma_{y} \to -\sigma_{y}$ and $\sigma_{z} \to - \sigma_{z}$.
This choice does not change the interaction term in the Hamiltonian, even in models
that account for the difference between intralayer and interlayer interactions.  It has the
advantage of implying the conservation of the $\hat{z}$ component of layer pseudospin
that is used to simplify our diagrammatic calculations. However, we do have to remember to invert the transformation
when we interpret the layer interchange interaction $\Gamma_{\rm X}$ which appears in our
calculations.  The inter-valley $\Gamma_{\rm X}$ scattering process is actually one
in which two electrons in the same layer are both scattered to the opposite layer.  It is of course
possible to perform our calculations without making this transformation simply by replacing
$\phi$ by $\pi-\phi$ in Eq.(\ref{eq:phasefactor}) for valley K'.
\end{thebibliography}
\end{document}